\documentclass[aps,prl,twocolumn, superscriptaddress, amsmath]{revtex4-1}
\usepackage{natbib}
\usepackage{graphicx}
\usepackage{bm}
\usepackage{color,soul}
\usepackage{hyperref}
\usepackage{float}
\usepackage{csquotes}
\usepackage{color,soul}
\usepackage{hyperref}
\hypersetup{colorlinks=true, urlcolor=blue, citecolor=blue}
\usepackage{longtable}
\usepackage{tabularx}
\usepackage{adjustbox}

\setcitestyle{journalcolor= blue}

\usepackage{babel}

\newcommand{\pzo}{{\rm PbZrO$_3$}}
\newcommand{\pzopt}{{\rm Pt/PbZrO$_3$/Pt }}

\begin{document}

\title{Ferroelectricity at the extreme thickness limit in the archetypal antiferroelectric \pzo}

\author{Nikhilesh Maity}
\email{nikhileshm@usf.edu}
\affiliation{Department of Physics, University of South Florida, Tampa, Florida 33620, USA}
\author{Milan Haddad}
\affiliation{School of Materials Science and Engineering Georgia Institute of Technology, Atlanta, GA 30318, USA}
\author{Nazanin Bassiri-Gharb}
\affiliation{G.W. Woodruff School of Mechanical Engineering Georgia Institute of Technology, Atlanta, GA 30318, USA}
\author{Amit Kumar}
\affiliation{Centre for Quantum Materials and Technologies, School of Mathematics and Physics, Queen's University Belfast, Belfast, UK}
\author{Lewys Jones}
\affiliation{School of Physics, Trinity College Dublin, Dublin, Ireland}
\affiliation{Advanced Microscopy Laboratory, Centre for Research on Adaptive Nanostructures and Nanodevices (CRANN), Dublin 2, Ireland}
\author{Sergey Lisenkov}
\affiliation{Department of Physics, University of South Florida, Tampa, Florida 33620, USA}
\author{Inna Ponomareva}
\email{iponomar@usf.edu}
\affiliation{Department of Physics, University of South Florida, Tampa, Florida 33620, USA}
\date{\today}

\begin{abstract}

Size-driven transition of an antiferroelectric into a polar ferroelectric or ferrielectric state is a strongly debated issue from both experimental and theoretical perspectives. While critical thickness limits for such transitions have been explored, a bottom-up approach in the ultrathin limit considering few atomic layers could provide insight into the mechanism of stabilization of the polar phases over the antipolar phase seen in bulk \pzo. Here, we use first-principles density functional theory to predict the stability of polar phases in \pzopt\ nanocapacitors. In a few atomic layer thick slabs of \pzo\ sandwiched between Pt electrodes, we find that the polar phase originating from the well established \textit{R3c} phase of bulk \pzo\ is energetically favorable over the antipolar phase originating from the \textit{Pbam} phase of bulk \pzo. The famous triple-well potential of antiferroelectric \pzo\ is modified in the nanocapacitor limit in such a way as to swap the positions of the global and local minima, stabilizing the polar phase relative to the antipolar one. The size effect is decomposed into the contributions from dimensionality reduction, surface charge screening, and interfacial relaxation, which reveals that it is the creation of well-compensated interfaces that stabilizes the polar phases over the antipolar ones in nanoscale \pzo.

\end{abstract}

\maketitle

Antiferroelectrics are defined as materials that exhibit an antipolar phase in the absence of the electric field, which can be converted into a polar phase by the application of an external electric field~\cite{rabe2013antiferroelectricity}. Lead zirconate (\pzo) was the first antiferroelectric to be discovered and has been investigated since the 1950s~\cite{liu2018antiferroelectrics,randall2021antiferroelectrics}.  
For \pzo\, the antipolar phase has \textit{Pbam} symmetry, while the energetically competitive metastable polar phase is \textit{R3c}. The polar phase can be induced by the application of a strong enough electric field, and gives rise to the double hysteresis loops of polarization as a function of AC fields~\cite{pintilie2008coexistence,singh1995structure,kagimura2008first}. In addition to this well established polar phase, several other phases have been reported both experimentally~\cite{fesenko1978structural,wei2020unconventional}, and computationally~\cite{lisenkov2020prediction,aramberri2021possibility}. Among others, an \textit{Ima2} phase in which both polar and antipolar orderings co-exist was proposed by Aramberri et al. \cite{aramberri2021possibility}, and recent experimental reports support ferrielectric behavior in \pzo\ thin films, consistent with co-presence of polar and antipolar arrangements in this material and extremely energetically close polar and antipolar phases~\cite{yao2023ferrielectricity,milesi2021critical}. 

\begin{figure*}[t!]
\centering
\includegraphics[width=1\textwidth]{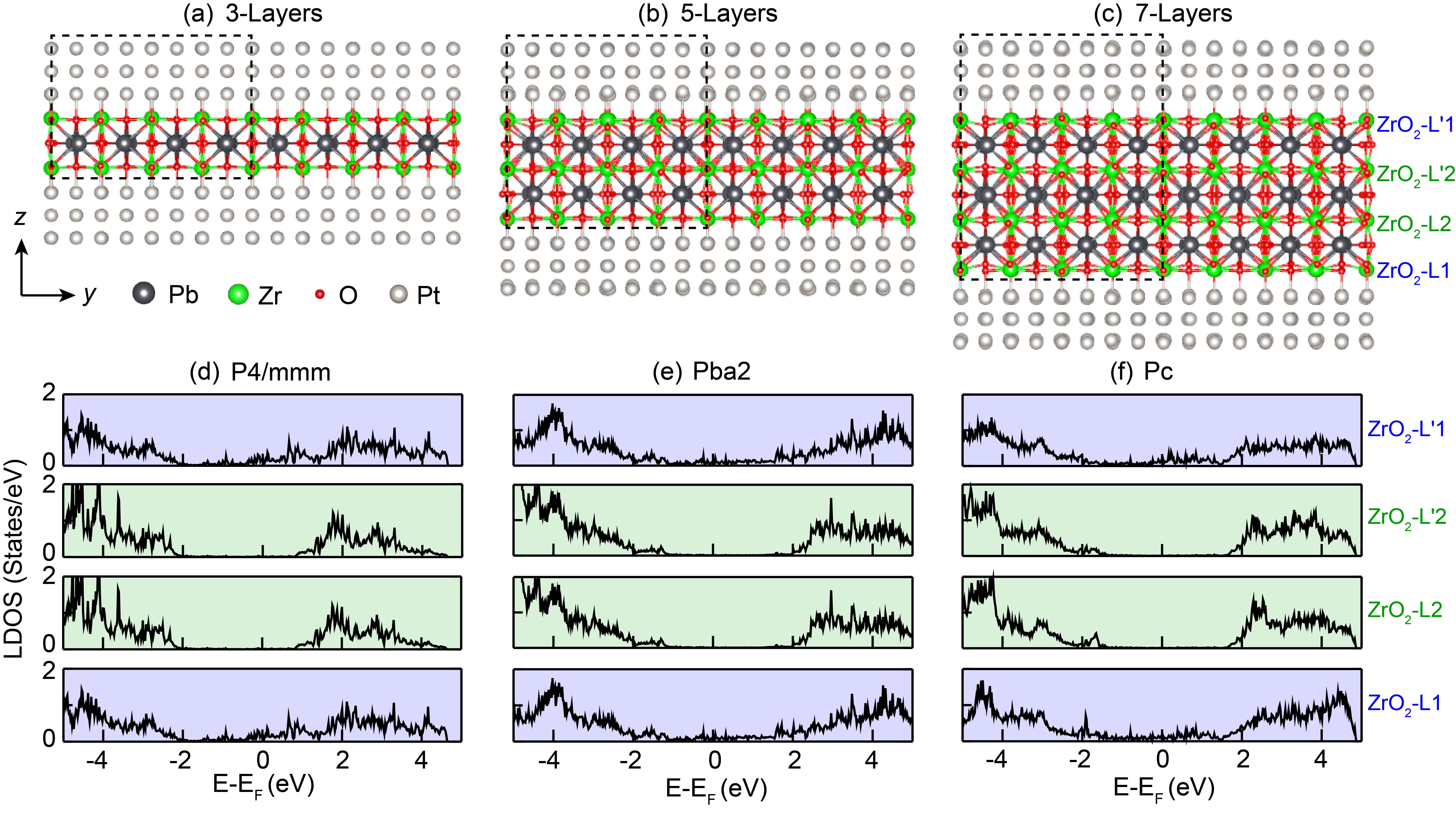}
\caption{\pzopt\ nanocapacitors with (a) three; (b) five; and (c) seven atomic layers of \pzo, sandwiched between three layers of cubic Pt. The dashed box indicates the simulation supercell, which is subjected to periodic boundary conditions in all three directions. Layer-by-layer LDOS of the ZrO$_2$ layers in the 7-Layer nanocapacitor for (d) \textit{P4/mmm}, (e) \textit{Pba2}, and (f) \textit{Pc} phases. The L1 and L$'$1 label layers next to  Pt, and the L2 and L$'$2 label the middle layers. Note, that for the middle layers the DOS at the Fermi level is smaller than our computational resolution.}
\label{fig1}
\end{figure*}

Indeed, upon scaling down, \pzo\ could exhibit rather different behavior.   Several experimental groups have reported ferroelectric or mixed ferroelectric/antiferroelectric behavior in \pzo\ films~\cite{pintilie2008coexistence,guo2019giant}. Chaudhuri et al. reported pinched hysteresis loop for a 22 nm epitaxial  \pzo\ film and rhombohedral ferroelectric phase for ultra thin films ($\leq$10~nm)\cite{chaudhuri2011epitaxial}. Likewise, transition into the ferroelectric phase has been reported in epitaxial antiferroelectric/ferroelectric multilayers of \pzo/Pb(Zr$_{0.8}$Ti$_{0.2}$)O$_3$ when individual layer thickness was below 10~nm~\cite{boldyreva2007thickness}. First-principles-based simulations have been used previously to explain the size-driven transition into ferroelectric phase~\cite{mani2015critical}. It was proposed that the presence of the surface favors the ferroelectric phase as it effectively removes the energetically costly interactions between head-to-tail dipoles. The same surface effect was found responsible for stabilization of ferroelectric phase in \pzo\, nanowires and nanodots~\cite{mani2016emergence}. More recently, weak ferroelectricity and a strong second harmonic generation (SHG) signal were observed in  110~nm   \pzo\ thin films at room temperature, suggesting that the room-temperature orthorhombic structure is non-centrosymmetric \cite{doi:10.1021/acsami.2c14291}. The polarization vector in the 45 nm \pzo\ film exhibited the magnitude and angle modulation modes, inducing quasi-antiferroelectric properties. Electron microscopy studies revealed that for 12 nm \pzo\ film, the polarization vector is in-plane, while for 8.8 nm thick film, the polarization vector has a large out-of-plane component~\cite{qiao2022polarization}. Transition into a ferroelectric-like state present in a large temperature window has been also reported in a 45~nm thick \pzo\ film~\cite{dufour2023ferroelectric}. Ferrielectric phase has been observed in the absence of external electric fields in highly oriented \pzo\ thin films, with modulations of amplitude and direction of the spontaneous polarization and large anisotropy for critical electric fields required for phase transition, calling for re-evaluation of fundamental science of antiferroelectricity~\cite{yao2023ferrielectricity}.

While these studies highlight different aspects and even some controversies associated with scaling down prototypical antiferroelectric \pzo, a knowledge gap exists in terms of investigating a bottom-up approach where atomically thin layers of \pzo\ (starting from a single unit cell consideration) give way to increasing number of layers. How does the interplay between antiferroelectricity, ferroelectricity, and ferrielectricity in \pzo\, manifest in just a few atomic layers of materials? Can
this interplay be understood from a bottom-up approach? Can scaling-up provide a different or complementary picture of these complex phenomena? The goals of this work are (i) to predict phases that can develop in atomically thin \pzopt nanocapacitors; (ii) to evaluate the possibility that contrary to the case of bulk \pzo, the ground state of such nanocapacitors can be polar; and (iii) to reveal the origin of such polar phase stabilization.

\begin{figure*}
\centering
\includegraphics[width=1\textwidth]{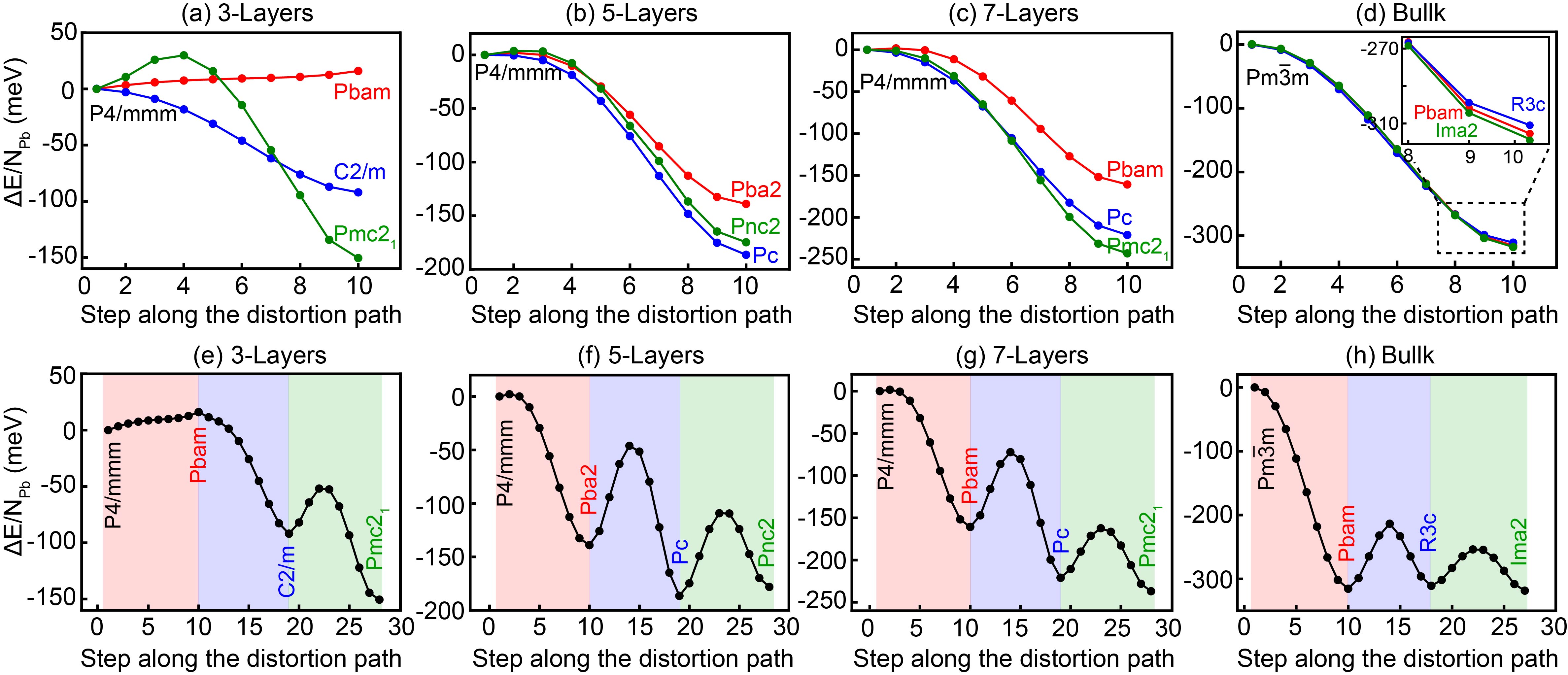}
\caption{Relative energy per Pb atom along the distortion path between different phases of \pzo\ heterostructures and bulk \pzo.}
\label{fig2}
\end{figure*}  
\section{Results}

Figure~\ref{fig2}(d) gives the energy evolution along the distortion path from cubic \textit{Pm$\bar{3}$m} bulk \pzo\ to the three low symmetry phases: antipolar \textit{Pbam}, polar \textit{R3c} and (multi)polar \textit{Ima2} \cite{aramberri2021possibility}. Note that we term \textit{Ima2} phase as multipolar as it exhibits both polar and antipolar orderings. At zero Kelvin, \textit{Pm$\bar{3}$m} phase is a local maximum, while the other phases are the (local) minima. Since the differences in energies are relatively small, it is important to eliminate the possibility of computational resolution affecting the results. More precisely, the size of computational supercells is different for different phases of \pzo, which could affect the computational resolution. To eliminate that effect, we created the distortion path that connects all four phases of this material and computed the energy evolution along such a path. The data are given in Figure~\ref{fig2}(h) and validate the relative stability of the phases. The associated energies of the phases are given in Table~\ref{table1}. The data predict that for bulk \pzo, the lowest energy phase is \textit{Ima2}, followed by \textit{Pbam} and \textit{R3c}. This is in agreement with the previous finding of Ref.~\cite{aramberri2021possibility}. However, the energy difference between \textit{Ima2} and \textit{Pbam} is only 2.3 meV, and Ref.~\cite{aramberri2021possibility} also alerts that this prediction is dependent on the exchange correlation functional. Given the limited experimental confirmation of the existence of the \textit{Ima2} phase but rather structures possibly comparable with it, we will consider this phase with caution. However, we do note that a ferrielectric phase would be most compatible with a body of recent experimental works that have cast doubt of the antiferroelectric nature of \pzo~\cite{an2023tuning,roleder2023weak,yao2023ferrielectricity,PhysRevLett.130.216801}.

\begin{table*}
\caption{Phases, relative energies, and polarizations for bulk \pzo\, and its heterostructures.}
\begin{adjustbox}{width=1.0\textwidth}
\begin{tabular}{| p{4.5em} | p{4.5em}  p{5em}  p{6em} | p{3.5em}  p{5em}  p{6em} | p{3.5em}  p{5em}  p{8em} | p{3.5em}  p{5em}  p{6.5em}|}
\hline

& Phase & $\frac{\Delta E}{N_{Pb}}$ (meV) & $\textbf{P}$ ($\mu$C/cm$^2$) & Phase & $\frac{\Delta E}{N_{Pb}}$ (meV) & $\textbf{P}$ ($\mu$C/cm$^2$) & Phase &  $\frac{\Delta E}{N_{Pb}}$ (meV) &  $\textbf{P}$ ($\mu$C/cm$^2$) & Phase & 
$\frac{\Delta E}{N_{Pb}}$ (meV) & $\textbf{P}$ ($\mu$C/cm$^2$) \\

\hline
$\textbf{Bulk}$              & \textit{Pm$\bar{3}$m} &  0.0   &  - & \textit{Pbam}  &  -315.4    &  -  & \textit{R3c}  &  -310.9 &  (33.7, 33.7, 33.7)  & \textit{Ima2}  &  -317.7 & (14.9, 0.0, 0.0)  \\
\hline
$\textbf{3-Layers}$          & \textit{P4/mmm}       &  0.0   &  - & \textit{Pbam}  &   15.9    &  -   & \textit{C2/m} &  -92.1  &  -    & \textit{Pmc2$_1$}  &  -133.8 & (20.5, 0.0, 0.0) \\
$\textbf{5-Layers}$          & \textit{P4/mmm}       &  0.0   &  - & \textit{Pba2}  &  -139.1   &  (0.0, 0.0, 4.9)  & \textit{Pc}  &  -186.5 &  (12.5, 14.6, 25.9)  & \textit{Pnc2}  &  -174.8 & (14.8, 0.0, 0.0) \\
$\textbf{7-Layers}$          & \textit{P4/mmm}       &  0.0   &  - & \textit{Pbam}  &  -160.8   &  -  & \textit{Pc}  &  -220.9 &  (22.1, 20.0, 26.6) & \textit{Pmc2$_1$}  &  -242.9 & (7.5, 0.0, 0.0) \\                       
\hline
\end{tabular}
\label{tab:mobility}
\end{adjustbox}
\label{table1}
\end{table*}

Next, we carry out the same energetic analysis for the nanocapacitors made of  3, 5, or 7 atomic layers of \pzo\  sandwiched between three atomic layers of cubic Pt (see Figure~\ref{fig1}(a)-(c)). Note that the thickness of Pt slab is limited by the computational cost. These are constructed by cutting bulk \pzo\ along [001] pseudocubic direction. Each one of the four \pzo\ phases described above was used for this construction to yield four initial phases for each nanocapacitor. For example, for the nanocapacitor with 3 atomic layers (to be termed as 3-Layers), we created four initial phases from \textit{Pm$\bar{3}$m}, \textit{Pbam}, \textit{R3c}, and \textit{Ima2}. The \pzo\ was simulated with ZrO$_2$ termination based on the experimental evidence, given the known high volatility of Pb compared to Zr, resulting often in Pb-deficient surfaces in Pb-based perovskites~\cite{gueye2017electrical,hardtl1969pbo}. Such nanocapacitors are then subjected to structural relaxation. We constrain in-plane lattice constant of the nanocapacitor to the lattice parameters of the given bulk phase of \pzo\ in order to model relaxed \pzo\ films. The lattice parameter of the nanocapacitor along the growth direction is relaxed. All ionic positions are fully relaxed. The \pzo\ phases produced in such relaxations are summarized in Table~\ref{table1}. All structures are provided in Ref.~\cite{ourgithub}.

It has been previously shown that underestimation of the band gap by DFT may lead to electron spillage in insulator-metal heterostructures, which leads to the artificial potential difference and erroneous predictions~\cite{stengel2011band}. To test whether  \pzopt\ nanocapacitors present such ``pathological regime", we computed local density of states (LDOS) for them and present these data in Figure~\ref{fig1}. The data show that no density of states appears at the Fermi level, which is an indication that the nanocapacitors are not in the pathological regime~\cite{stengel2011band}.

Figure~\ref{fig2} shows the energy evolution along the distortion path for the nanocapacitors. The energy landscape is quite different from the bulk one. Most strikingly, the phase that originated from the antipolar \textit{Pbam} phase is now higher in energy than the ones that originated from both polar phases (\textit{R3c} and \textit{Ima2}). This is in agreement with earlier reports of ferroelectric phase stabilization in ultrathin \pzo\ films obtained from the effective Hamiltonian simulations~\cite{mani2015critical}. We have computed polarization along the associated distortion paths (see Figure S2 of Supplementary material) and list the values in Table~\ref{table1}. Technically, the polarization was computed within Berry phase approach by removing Pt and decreasing c-lattice parameter to simulate periodic structure.  We also computed the local polarizations in the polar phases of the nanocapacitors and found that the $P_z$ values are slightly smaller in the interface layers (see Figure S8 in Supplementary Materials).  The out-of-plane polarization component  gradually from bulk to 5-Layers slab and disappears in 3-Layers one. We attribute that to the increased role of the depolarizing field~\cite{junquera2003critical} and structural changes due to surface relaxation. In particular, the depolarizing field is a function of the effective screening length, $\lambda$, for an electrode~\cite{ghosez2006first}. For example, $E_{dep}=-\frac{-8\pi P \lambda}{d}$, where $P$ is the polarization and $d$ is the slab thickness \cite{junquera2003critical}. The relationship explains why depolarizing field increases as the thickness of the slab decreases, therefore, leading to suppression of polarization. Indeed, we found in computations that residual depolarizing field is larger in 5 layer slab as compared to 7 layer one (see Supplemental Materials). Note, that we have also computed the energy landscapes as in Figure~\ref{fig2}(a)-(d) using PBEsol functional and find the same qualitative dependencies (see Figure S7 of Supplementary Materials).

\begin{figure*}
\centering
\includegraphics[width=1\textwidth]{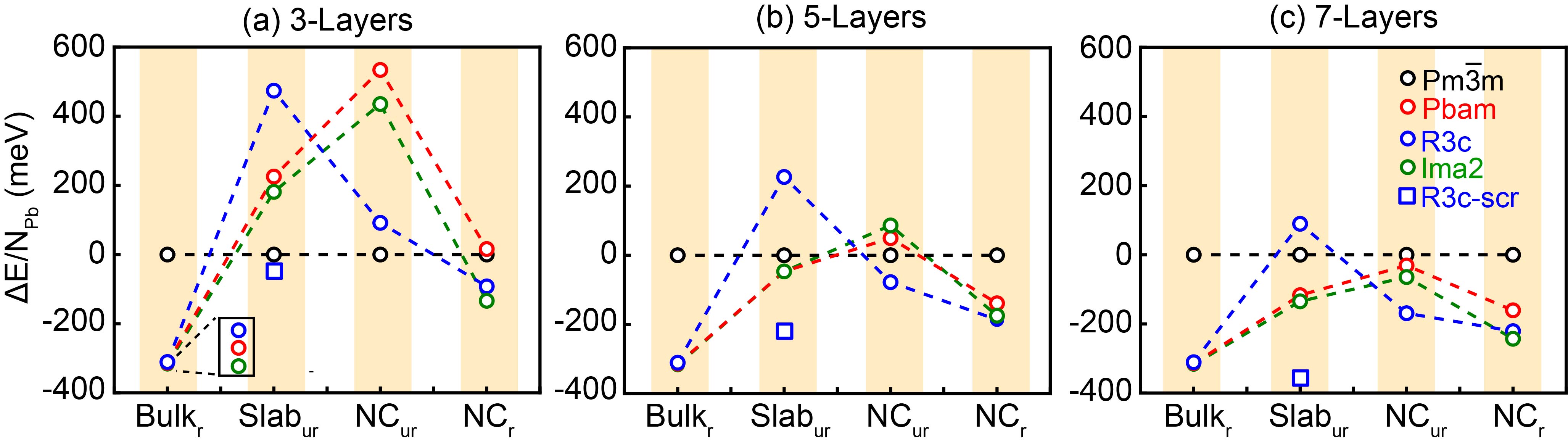}
\caption{Relative energy per Pb atom in bulk \pzo\ (Bulk$_r$) and  \pzopt\  of different types: unrelaxed slab without Pt (Slab$_{ur}$); unrelaxed nanocapacitor (NC$_{ur}$); relaxed nanocapacitor (NC$_r$). Note, that the phases are labeled by the bulk phase they originate from to facilitate the comparison between different nanocapacitors/slabs.  Table \ref{table1} should be used if the space group of a given nanocapacitor is needed. The blue squares give the relative energies of the fully screened slabs as described in the text.}
\label{fig3}
\end{figure*}  

Thus, our computational data suggest that in ultrathin slabs of 5 and 7 \pzo\ layers sandwiched between Pt electrodes, the polar phases with significant polarization are stabilized. To confirm the stability of the polar phase in 5 Layer thick nanocapacitor we carried out \textit {ab initio} molecular dynamics simulations at 3~K (see Figure S3 of Supplementary Materials). The phase was found  to be stable and  lower in energy than the antipolar one. Previously, several mechanisms of stabilizing polar \textit{R3c} phase in orthorhombic  CaTiO$_3$-type have been proposed.  Strain and electrical boundary conditions were predicted as a mean to stabilize metastable polar phase   under conditions of relatively high strain and electric displacement fields \cite{PhysRevLett.128.197601}. However, our layers are stress-free and are likely to be under nearly zero electric displacement field, thanks to the screening from the electrodes.  Another approach to stabilizing polar phase in CaTiO$_3$-type structure is through oxygen octahedra rotation engineering in perovskite superlattices \cite{PhysRevX.6.011027}. However, this mechanism is not activated in our layers as they do not experience epitaxial boundary conditions.
To elucidate the origin of polar phase stabilization in our case   we  take the advantage of first-principles simulations to decouple the size effects in \pzo\ nanocapacitor into contributions from (i) dimensionality reduction (creation of slabs from bulk); (ii) electrostatic boundary conditions (surface charge screening); and (iii) interfacial relaxation. To accomplish that, we start with the four fully relaxed phases of bulk \pzo\ (\textit{Pm$\bar{3}$m}, \textit{Pbam}, \textit{R3c}, and \textit{Ima2}). Their energy with respect to the cubic phase is given in Figure~\ref{fig3} (labeled as relaxed bulk: Bulk$_{r}$). The relative energetic ordering of the phases is the same as already described. Next, we cut 3, 5, and 7 Layers slabs along [001] pseudocubic direction for each of the phases. To simulate open-circuit boundary conditions, we consider supercell of \pzo\ slabs separated by 8 \AA\ vacuum. This value of vacuum provides convergent results as has been verified by increasing the value to 10 and 12 \AA. Details can be found in Supplementary Materials. No structural relaxation is carried out, which allows us to isolate the effect of dimensionality reduction. The associated energy is given in Figure~\ref{fig3} (labeled as unrelaxed slab: Slab$_{ur}$). Note that to aid the comparison, we label the phases by the space group of bulk from which they originate. The actual space groups could differ. For all slabs, we find that \textit{R3c}-originated phase is the highest in energy. To understand this finding we recall that when the phase has out-of-plane polarization component it is  severely penalized by the strong depolarizing field associated with the direction of reduced dimensionality. Such energy penalty can be estimated as $V\frac{P_z^2}{2\varepsilon_\infty\varepsilon_0}$, where $V$ and $P_z$ are the volume  and  the out-of-plane polarization component of the slab respectively, while $\varepsilon_\infty$ is optical dielectric constant. The squares in Figure~\ref{fig3}  shows the energy values obtained after subtracting this contribution, which is equivalent of perfect surface charge screening. This results in the phases to be energetically most favorable. With the long-range electrostatic effects out of equation, the difference in the stability of the phases with respect to the one in bulk can be attributed to the surface energy, that is the energetic cost associated with creation of a surface. We computed the surface energy (see Supplementary Materials) for the different phases, which confirmed that  \textit{R3c}-originated phases  have lowest  surface energy, when depolarizing field is fully screened.

A noteworthy finding is that for 3-Layers slab, the energy of the cubic-originated phase is now lower than the \textit{Pbam} and \textit{Ima2}-originated phases. This could be explained by the fact that such slab has only single oxygen octahedron along the growth direction. Cooperative oxygen octahedra tilts are known to stabilize lower symmetry phases in \pzo. As such, tilts can no longer cooperate along the growth direction in a 3-Layers slab, and the lower symmetry phases do not stabilize with respect to cubic one in this case. This is further supported by the data for 5- and 7-Layers where such cooperation is possible, and the lower symmetry phases are now lower in energy with respect to cubic-originated one (except for the \textit{R3c}-originated phase, of course). Next, we turn on surface charge screening by sandwiching the slabs from the previous calculations between Pt electrodes, which are expected to provide good surface charge compensation. The interfacial distance between Pt and ZrO$_2$ layers is taken to be 2 \AA. Again, we do not allow any structural relaxations at this point to isolate the effect of screening. The relative energies for such nanocapacitors are given in Figure~\ref{fig3} (labeled as nanocapacitor unrelaxed: NC$_{ur}$). As expected, surface charge screening lowered the energy of the \textit{R3c}-originated phase. The striking result is that now, for all nanocapacitors considered, the \textit{R3c}-originated phase is lower in energy than \textit{Pbam} and \textit{Ima2}-originated phases. For 5- and 7-Layers, the phase is polar. Thus, we find that for well compensated \pzo\ ultrathin slabs, the polar phase is more stable than the antipolar phase. Note, that the energy of the \textit{R3c}-originated phase in the nanocapacitor is higher than the one of the fully screened slab. This is due to the fact that Pt electrodes do not provide complete surface charge screening, as previously argued. The last contribution to be quantified is the effect of atomic relaxation at the interface between \pzo\ and Pt. This can be accomplished by subjecting the nanocapacitors of the previous step to full structural relaxation, which was already done in producing data for Figure~\ref{fig2}. Here, we just augment Figure~\ref{fig3} data with those energies (labeled as nanocapacitors relaxed: NC$_{r}$). From the Figure, we can see that in all cases, interfacial relaxation (i) stabilizes the lower symmetry phases over the cubic-originated one; (ii) does not change the relative ordering of the polar and antipolar phases (the polar phases remain lower in energy than the antipolar one). The dipole patterns due to polar and antipolar modes in different nanocapacitors are given in  Figure S4.

Thus, our analysis reveals that reducing dimensionality of \pzo\ from 3D (bulk) to 2D (slabs) activates size effects that stabilize polar phases over the antipolar ones. In particular, the surface energy is phase-dependent and is lower for the phases that originate from the \textit{R3c} of bulk. However, this effect could be overpowered by the depolarizing field, favoring phases with small or zero out-of-plane polarization. Therefore, for the cases of good surface charge compensation, we expect polar phases to be stabilized in nanocapacitors with ultrathin \pzo. Pt electrodes are found to provide such good compensation except for the thinnest slab of 3 atomic layers. Indeed, we find only negligible residual depolarizing field in fully relaxed heterostructures (see Supplementary Materials).

Interestingly, we find the polar \textit{Pba2} phase in a 5-Layers nanocapacitor. This is actually a consequence of the symmetry. In bulk \textit{Pbam} phase of \pzo\, the mirror plane $m$ is the PbO plane. As can be seen from Figure~\ref{fig1} such mirror plane can be preserved in slabs with odd number of PbO layers, but is lost in slabs with even number of PbO layers, consequently lowering the symmetry of the phase. The polar axis gives origin to the tiny polarization of 4.9 $\mu$C/cm$^2$. Consequently, we expect this effect to be present in all nanocapacitors with an even number of Pb layers. Note, that layer-by-layer decomposition of polarization (Fig. S8(c)) indicates that the local polarization is present in all layers, not just the interface ones. 

Are the nanocapacitors ferroelectric? By definition, the direction of polarization needs to be reversible by the application of electric field. From Figure \ref{fig2}, we find that the energy barriers between different phases of nanocapacitors are comparable to those of bulk \pzo. Since the reversibility of polarization direction in bulk \pzo\ is well established experimentally, we conclude that these barriers are surmountable, and the polarization direction in nanocapacitors can be reversed by experimentally accessible electric fields. In the context of perovskite ferroelectrics, there has been a long running debate about the thickness limit below which the material might not be able to sustain polarization \cite{junquera2003critical}. The results obtained in this work suggest that scaling up an antiferroelectric layer-by-layer might offer access to genuinely polar phases at the ultrathin limit. 

In summary, we have used DFT simulations to investigate relative phase stability of polar, antipolar, and multipolar phases of \pzopt\ nanocapacitors. The energy landscape of bulk \pzo\ consists of the local maximum associated with \textit{Pm$\bar{3}$m} cubic phase and local minima associated with \textit{Pbam}, \textit{R3c}, and \textit{Ima2}  phases. Among experimentally observed phases, \textit{Pbam} is the global minimum, while \textit{R3c} is a local one, as well established for this prototypical antiferroelectric. In nanocapacitors of 3-, 5- and 7-atomic layer \pzo\ sandwiched between Pt, the energetic ordering is reversed, and the polar phases are associated with minima, which are lower in energy than the antipolar ones, thus revealing that such heterostructures are ferroelectric. To understand the origin of the behavior opposite to the bulk one, we used computations to isolate energy contributions to different phases due to dimensionality reduction, surface charge compensation, and interfacial relaxation. It was found that the combination of reduced dimensionality and good surface charge compensation provided by  Pt is responsible for the stabilization of polar phases with respect to the antipolar one in \pzopt\ nanocapacitor. We also report that the out-of-plane polarization component reduces with \pzo\ thickness, which is attributed to the depolarization effects. However, the in-plane polarization component increases in some cases. We believe that our findings shed light onto size effects in this prototypical antiferroelectric and will be valuable in interpreting diverse experimental data. They could also stimulate untraditional ways to achieve ferroelectricity at the extreme thickness limits.

\section{Methods}

The first-principles density functional theory (DFT) calculations were performed using Vienna ab initio simulation package (VASP)~\cite{kresse1996efficient,kresse1996efficiency}. The all-electron projector augmented wave (PAW) potentials~\cite{blochl1994projector,kresse1999ultrasoft} were used to represent the ion-electron interactions. For the calculations, the electronic exchange and correlation part of the potential was described by the local density approximation (LDA)~\cite{ceperley1980ground}. The choice of the exchange correlation functional has been motivated by its performance on ferroelectric \cite{PhysRevMaterials.4.073802} and antiferroelectric (\pzo) \cite{aramberri2021possibility} perovskites. In particular, a thorough investigation on the performance of LDA, PBE, PBEsol and SCAN functional  in predicting relative phase stability of \pzo\ phases, including at finite temperatures, reported that predictions from SCAN and PBE cannot be reconciled with experimental data, while LDA  and PBEsol provide qualitatively the same predictions \cite{aramberri2021possibility}. Likewise, for ferroelectric perovskites  LDA  and PBEsol were identified as the best performing functionals \cite{kresse1996efficient,kresse1996efficiency}. The Kohn-Sham orbitals were expanded using the plane wave basis sets with an energy cutoff of value 600 eV. All structures were relaxed using the conjugate-gradient algorithm until the Hellmann-Feynman forces on every atom are less than 0.005 eVÅ$^{-1}$. The $\Gamma$-centered k-point mesh with the spacing of 0.22 \AA$^{-1}$ was used in our calculations. The polarization of the systems was calculated according to the modern theory of polarization~\cite{king1993theory,vanderbilt1993electric,resta1994macroscopic}. The space groups were identified, and distortion paths were constructed using the ISOTROPY suite~\cite{campbell2006isodisplace,stokes2005findsym}. The ISOTROPY defines the distortion path as linear interpolation between its end points. We used tolerances 0.1 \AA~ for both the lattice and atomic positions. 

\section{Acknowledgements}
N.M and S.L acknowledge financial support by the U.S. National Science Foundation under grant No.DMR-2219476.  I.P. acknowledges financial support by the U.S. Department of Energy, Office of Basic Energy Sciences, Division of Materials Sciences and Engineering under grant DE-SC0005245. Computational support was provided by the National Energy Research Scientific Computing Center (NERSC), a U.S. Department of Energy, Office of Science User Facility located at Lawrence Berkeley National Laboratory, operated under Contract No. DE-AC02-05CH11231 using NERSC award BES-ERCAP-0025236. L.J. acknowledges support from SFI grant SFI/21/US/3785. A.K. gratefully acknowledges support from Department of Education and Learning NI through grant USI-211.

\newpage

\onecolumngrid
\begin{center}
   \textbf{\Large Supplementary Material}
\end{center}

\renewcommand{\thefigure}{S\arabic{figure}}
\setcounter{figure}{0}

\begin{figure}[h!]
\centering
\includegraphics[width=1.0\textwidth]{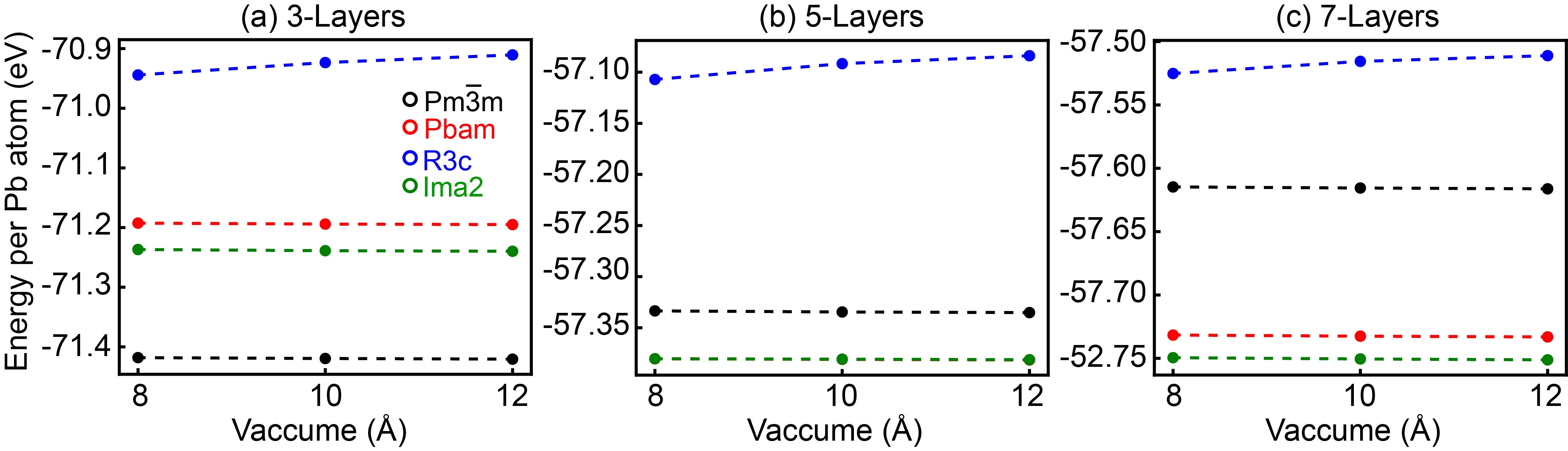}
\caption{The energy convergence test of vacuum in the slab geometry with respect to the thickness for different phases of \pzo.}
\label{FigS1}
\end{figure}

\begin{figure}[h]
\centering
\includegraphics[width=1.0\textwidth]{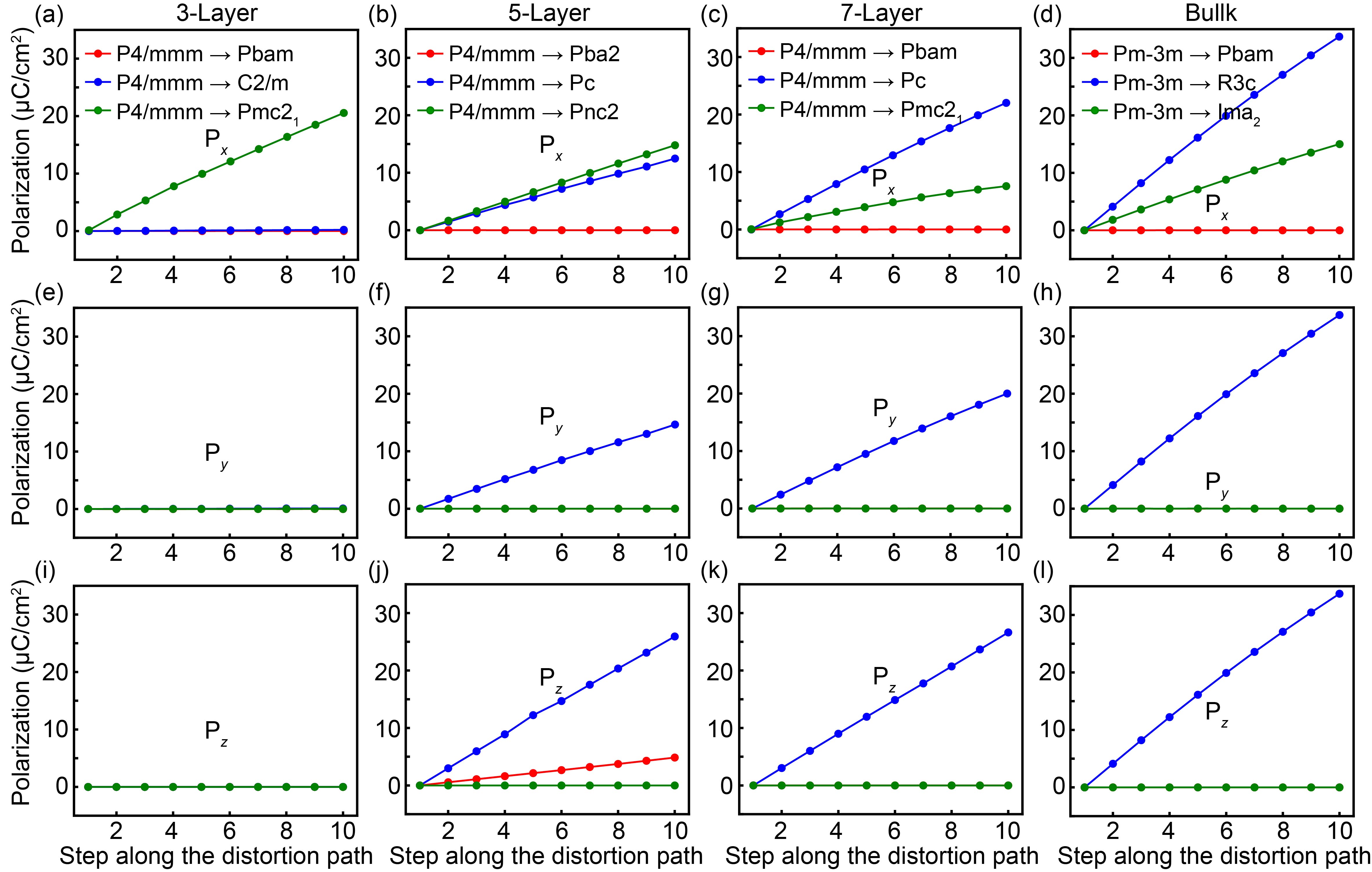}
\caption{Polarization along the distortion path between different phases of \pzo\ nanocapacitors and bulk \pzo.}
\label{FigS2}
\end{figure}

\begin{figure}[h]
\centering
\includegraphics[width=0.9\textwidth]{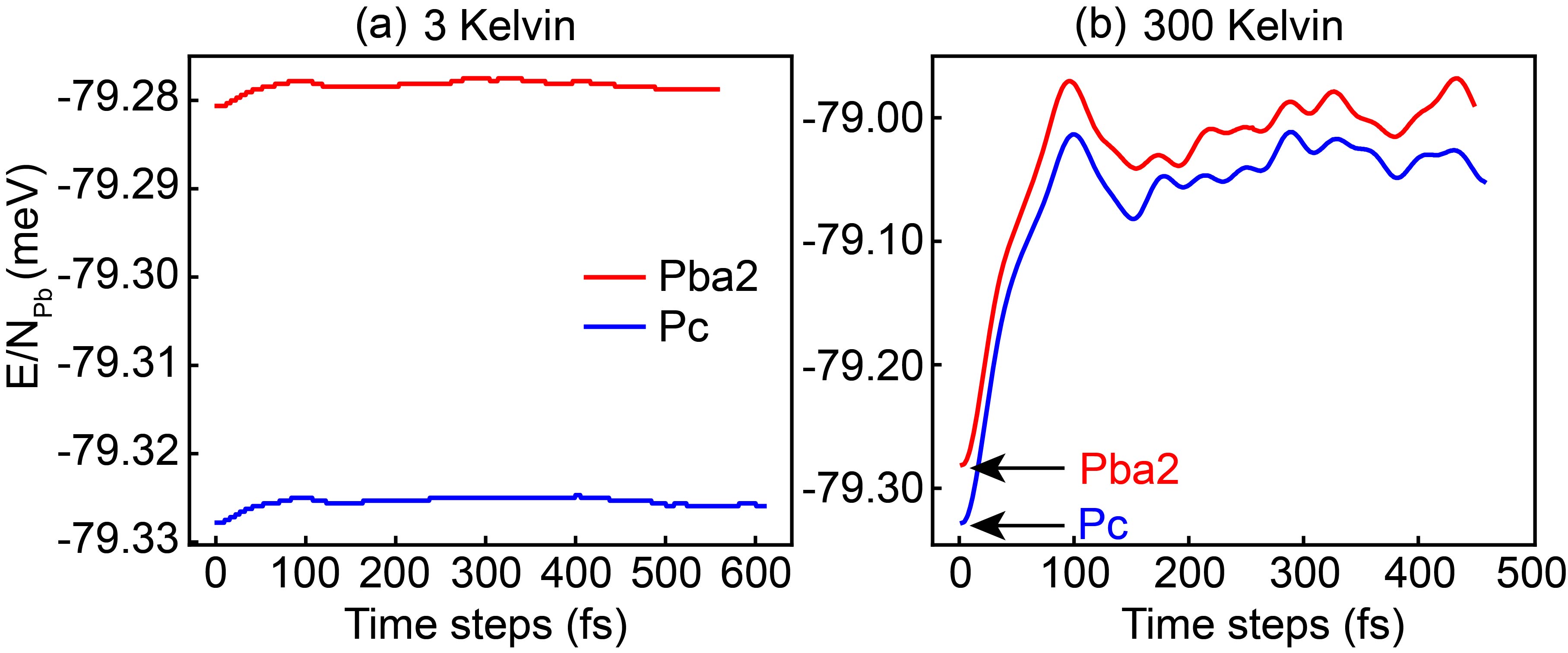}
\caption{\textit{Ab initio} molecular dynamics simulations for 5-Layer film of Pba2 and Pc phase for temperature (a) 3 Kelvin and (b) 300 Kelvin.}
\label{FigS3}
\end{figure}

\begin{figure}[h]
\centering
\includegraphics[width=1.0\textwidth]{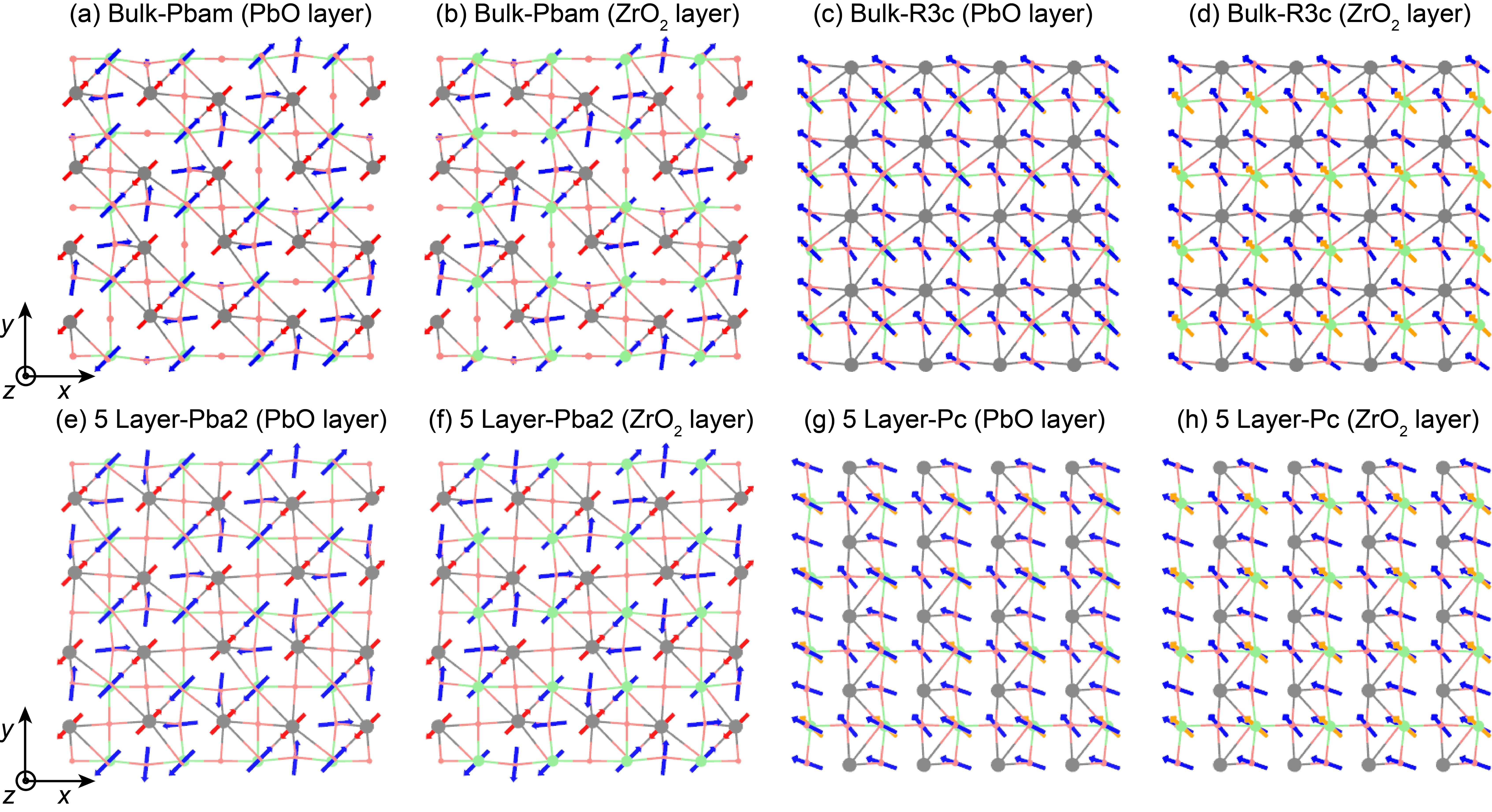}
\caption{Top view of the atomic displacements for (a)-(d) bulk and (e)-(h) 5-Layer nanocapacitor for  polar and antipolar modes. The phases are given in the titles.}
\label{FigS4}
\end{figure}

We have carried out convergence study with respect to the thickness of vacuum in the slab geometry. The dependence of the slab energy on the thickness of vacuum layer is given in Figure~\ref{FigS1}. The values are well converged with respect to the vacuum thickness.

\newpage

\section{Surface energy}
The surface energy of slab, $\sigma$, can be computed as 

\begin{equation}
\sigma = \frac{1}{2A} (E_{slab} - \mu_{Pb}N_{Pb} - \mu_{Zr}N_{Zr} - \mu_{O}N_{O})  
\end{equation}

where, $E_{slab}$ is the total energy of the slab, and $\mu_{Pb}$, $\mu_{Zr}$ and $\mu_{O}$ are the chemical potentials of each Pb, Zr and O atom, respectively. $A$ is the surface area of the slab. In addition, $N_{Pb}$, $N_{Zr}$ and $N_{O}$ are the numbers of the corresponding atoms in the slabs. In equlibrium, the total chemical potential of the elemental Pb, Zr and O is with that of bulk $PbZrO_3$: 

\begin{equation}
\mu_{PbZrO_3} (Bulk) =  \mu_{Pb} + \mu_{Zr} + 3\mu_{O} 
\end{equation}

\begin{equation}
\mu_{O}  = \frac{\mu_{PbZrO_3} (Bulk) - \mu_{Pb} - \mu_{Zr}}{3}
\end{equation}

The elemental O chemical potential was modeled by the corresponding chemical potential in bulk $PbZrO_3$.

\begin{figure}[h]
\centering
\includegraphics[width=0.5\textwidth]{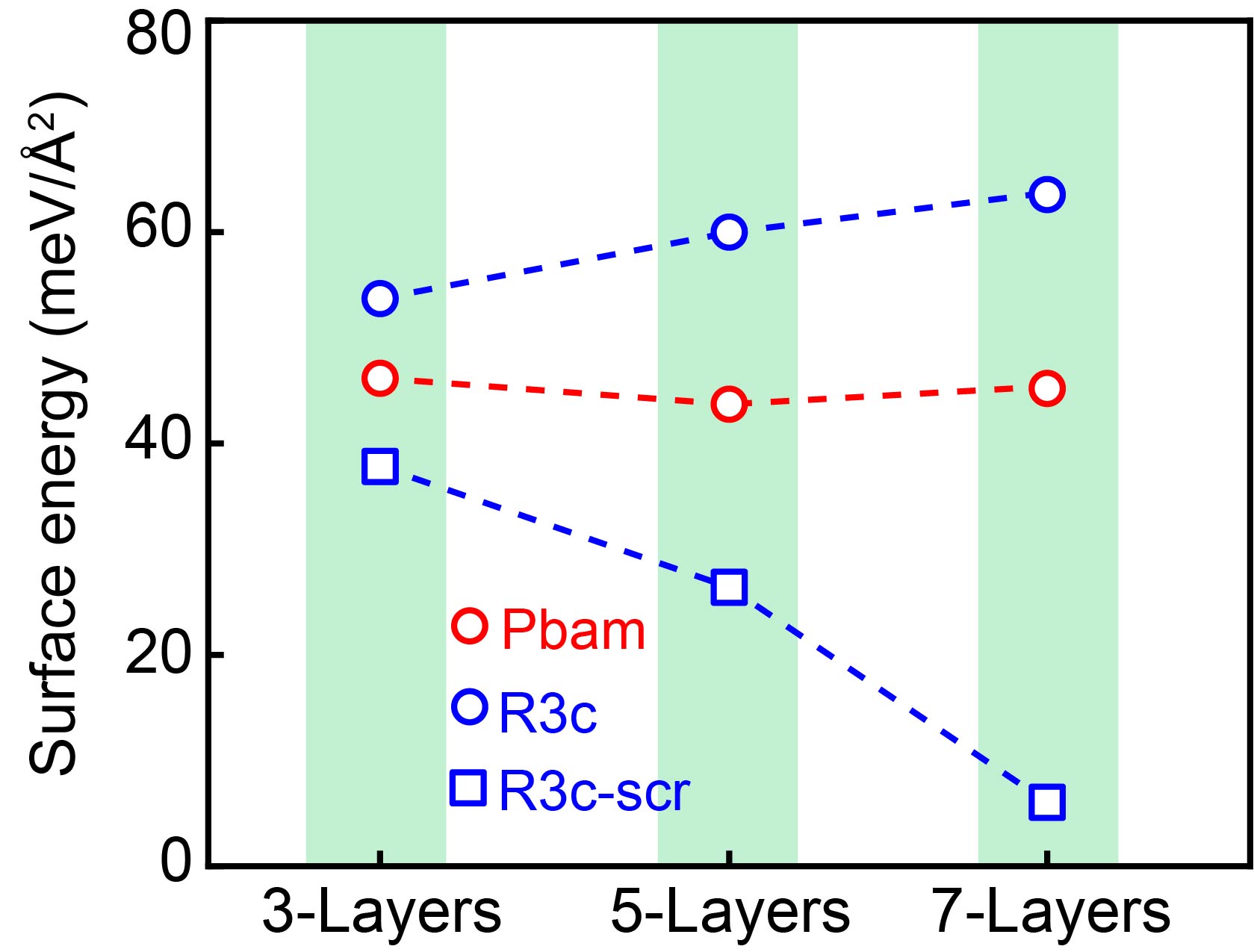}
\caption{Surface energy of $PbZrO_3$ slabs for different phases.}
\label{FigS5}
\end{figure}

Figure \ref{FigS5} shows the surface enenrgies of $Pbam$, $R3c$ and the $R3c$ phase with fully screened depolarizing  field (that is $R3c-scr$). We can see that the surface energy is lower for the fully screened $R3c$ phase as compared to $Pbam$.

\section{Residual depolarizing field}

\begin{figure}[h!]
\centering
\includegraphics[width=0.8\textwidth]{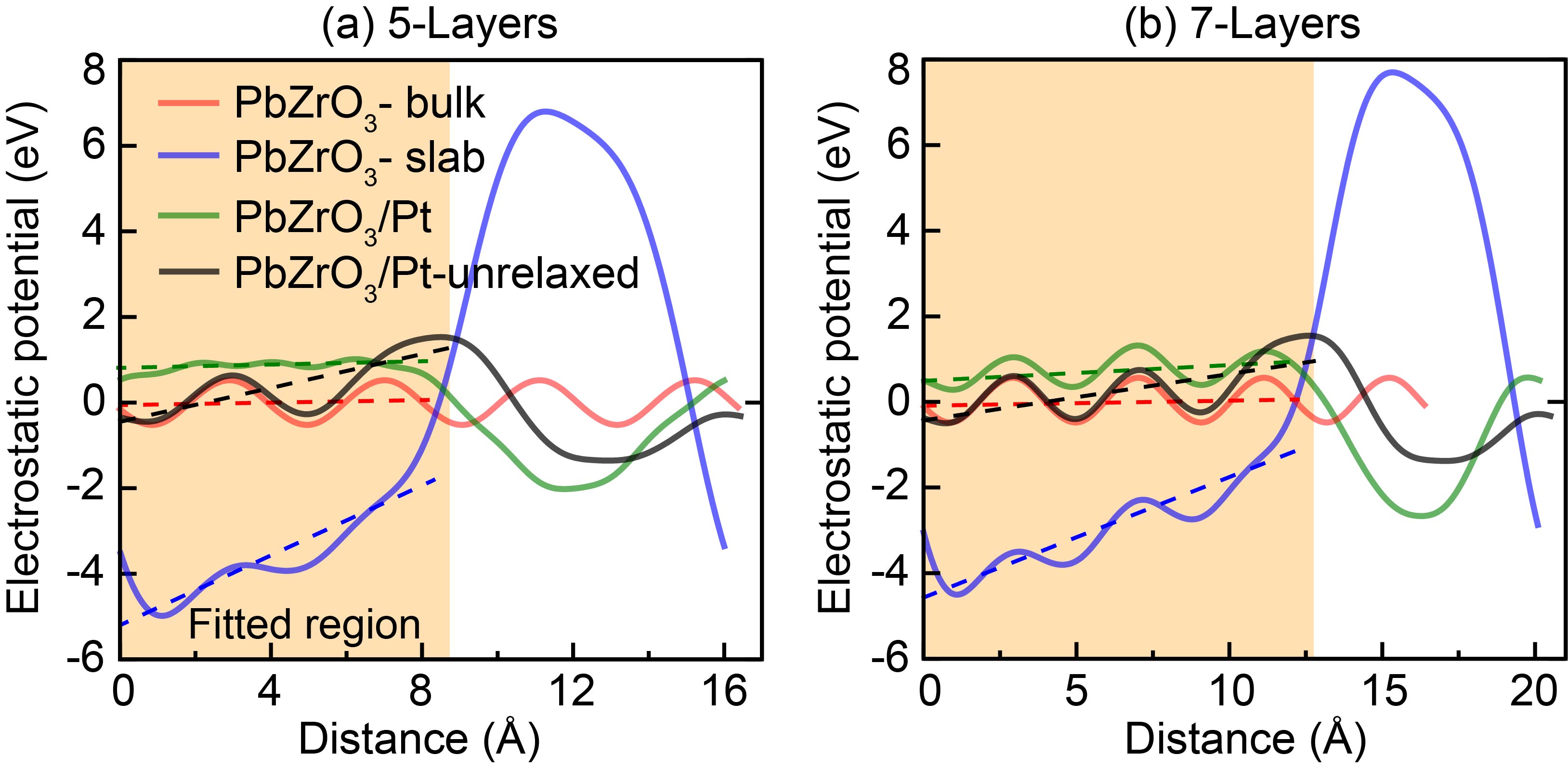}
\caption{Planar averaged electrostatic potential for the PbZrO$_3$ slabs separated by vacuum,  Pt/PbZrO$_3$/Pt heterostructure (both relaxed and unrelaxed) and bulk PbZrO$_3$.}
\label{FigS6}
\end{figure}

Figure~\ref{FigS6} shows the planar averaged electrostatic potential for the PbZrO$_3$ slabs separated by vacuum,  Pt/PbZrO$_3$/Pt heterostructure (both relaxed and unrelaxed) and bulk PbZrO$_3$. We can see no potential drop for bulk. The largest potential drop occurs for the slab geometry. The linear fit to the data gives 4.08$\pm$0.17~GV/m of the electric field for the case of 5 layers and 2.81$\pm$0.08~GV/m for the case of 7 layers.  These values are consistent with the maximum depolarizing field $E_{dep}=-P_z/\varepsilon_\infty\varepsilon_0=$5.0~GV/m expected for this geometry, where $\varepsilon_\infty=$7.0. Sandwiching the unrelaxed slab between Pt electrodes results in significant decrease of the slope and estimated residual depolarizing field of 1.97$\pm$0.13~GV/m and 1.08$\pm$0.08~GV/m for the 5 and 7 layers, respectively. Allowing the ions of PbZrO$_3$ to relax results in decrease of polarization in response to the  residual depolarizing field and consequential decrease in the field down to 0.19$\pm$0.03~GV/m and 0.38$\pm$0.05~GV/m values in 5 and 7 layers nanocacpacitors, respectively.   

\begin{figure}
\centering
\includegraphics[width=1\textwidth]{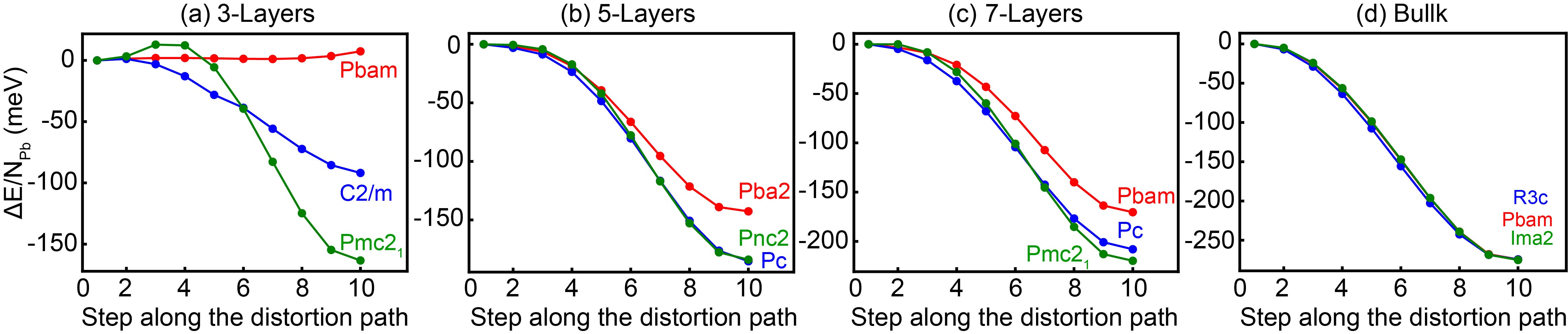}
\caption{Relative energy per Pb atom along the distortion path between different phases of \pzo\ heterostructures and bulk \pzo\, using PBEsol functional.}
\label{FigS7}
\end{figure}

\begin{figure}
\centering
\includegraphics[width=1\textwidth]{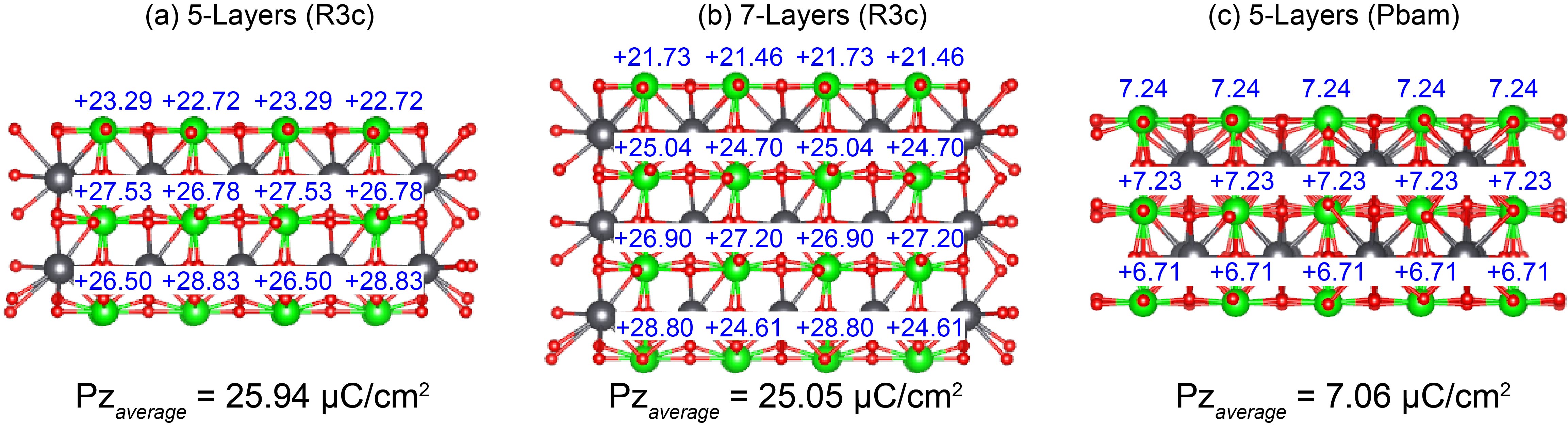}
\caption{The local out-of-plane polarizations (in $\mu$C/cm$^2$) for the polar phases of nanocapacitors: \textit{R3c} phase of 5-layer nanocapacitor (a) \textit{R3c} phase of 7-layer nanocapacitor (b) and \textit{Pbam} phase of 5-layer nanocapacitor (c).}
\label{FigS8}
\end{figure}

\section{Local polarization in nanocapacitors}

Figure~\ref{FigS8} shows the local polarization for  the polar phase of PbZrO$_3$ nanocapacitors with 5 and 7 layers. The values for each PbZrO$_3$ unit cell  were calculated as follows 
\begin{equation}
    \textbf{d}=\frac{1}{V}\sum_i Z^* (i) \textbf{u}(i)
\end{equation}
 where $V$ is the unit cell volume, $Z^* (i)$ is the Born effective charge of the ion $i$ and $\textbf{u}(i) $ is the displacement of ion $i$ from the cubic phase. The local polarization is centered on Zr ion since it has the largest displacement (see Figure S4). The summation goes over nearest neighbor Pb and O. The Born effective charges are +3.93, +5.92, $-$2.49, and $-$4.86 for Pb, Zr, and two nonequivalent O, respectively. For the surface unit cells we used $V=$ 62.20 \AA$^3$, which was computed using the volume per ion in PbZrO$_3$ unit cell. This method predicts local polarizations in \textit{R3c} phase of bulk PbZrO$_3$ of 37.22~$\mu$C/cm$^2$, which compares reasonable with the Berry phase value of supercell polarization of 33.7~$\mu$C/cm$^2$. 

\begin{figure}
\centering
\includegraphics[width=0.8\textwidth]{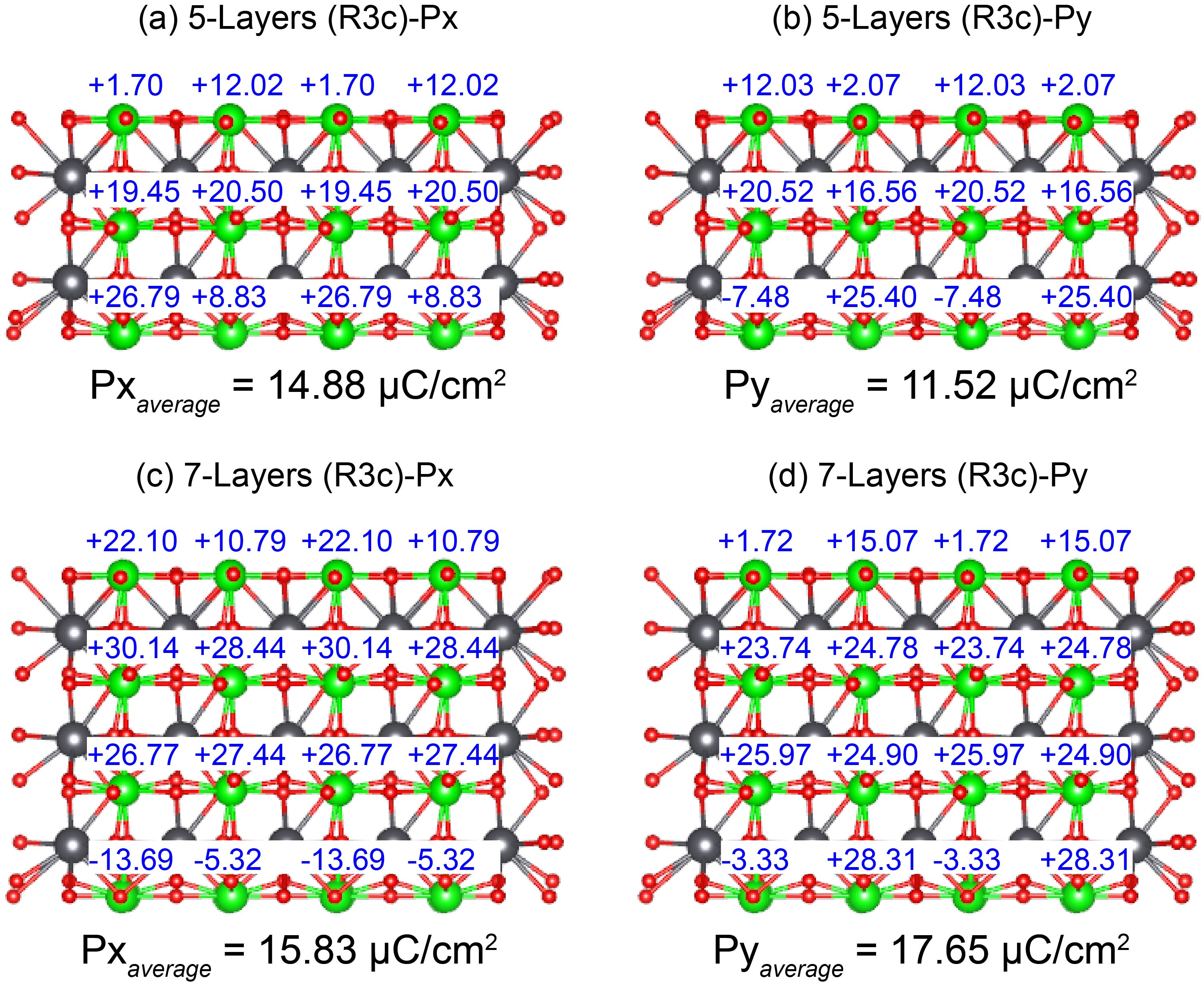}
\caption{The local in-plane polarizations (in $\mu$C/cm$^2$) for the polar phases of nanocapacitors: \textit{R3c} phase of 5-layer nanocapacitor (a-b) and \textit{R3c} phase of 7-layer nanocapacitor (c-d).}
\label{FigS9}
\end{figure}

Figure~\ref{FigS8} and Figure~\ref{FigS9} report the local out-of-plane and in-plane polarizations components in 5 and 7 layers nanocapacitors, respectively. We find the values to be slightly smaller in the interface layers. 

\newpage

\end{document}